\documentclass[11 pt]{article}
\usepackage[utf8]{inputenc}
\usepackage[english]{babel}
\usepackage{amsmath}
\usepackage{enumerate}
\usepackage{amsthm}
\usepackage{latexsym}
\usepackage{amssymb}
\usepackage[pdftex]{graphicx}
\usepackage{float}
\usepackage{textcomp}
\usepackage{appendix}
\usepackage{xcolor}
\theoremstyle{definition}

\newcommand{\mb}[1]{\mathbf{#1}}
\newcommand{\ms}[1]{\boldsymbol{#1}}
\newcommand{\dd}{\mathrm{d}}
\newcommand{\argmax}[2]{%
\smash{\mathop{{\rm argmax}}\limits_{#1}}\,#2} 

\def\pp{{\mathbb P}}

\usepackage[textwidth=16cm,footskip=1.5cm,textheight=23cm,marginratio=1:1,heightrounded]{geometry}

\author{Merlin Keller, Guillaume Damblin, Alberto Pasanisi, Mathieu Schumann, \\ Pierre Barbillon, Fabrizio Ruggeri, Eric Parent}
\title{Validation of a computer code for the energy consumption of a building, with application to optimal electric bill pricing}

\begin{document}

\maketitle

\begin{abstract}

%
%

In this paper, we propose a practical Bayesian framework for the calibration and validation of a computer code, and apply it to a case study concerning the energy consumption forecasting of a building. Validation allows to quantify forecasting uncertainties in view of the code's final use. Here we explore the situation where an energy provider promotes new energy contracts for residential buildings, tailored to each customer's needs, and including a guarantee of energy performance.

Based on power field measurements, collected from an experimental building cell over a certain time period, the code is calibrated, effectively reducing the epistemic uncertainty affecting some code parameters (here albedo, thermal bridge factor and convective coefficient). Validation is conducted by testing the goodness of fit of the code with respect to field measures, and then by propagating the {\em a posteriori} parametric uncertainty through the code, yielding probabilistic forecasts of the average electric power delivered inside the cell over a given time period. 

To illustrate the benefits of the proposed Bayesian validation framework, we address the decision problem for an energy supplier offering a new type of contract, wherein the customer pays a fixed fee chosen in advance, based on an overall energy consumption forecast. According to Bayesian decision theory, we show how to choose such a fee optimally from the point of view of the supplier, in order to balance short-terms benefits with customer loyalty.

\end{abstract}

\section{Introduction}

Today, more and more green buildings are designed in order to maintain a low energy consumption with the objective of dividing by four the CO$2$ emissions in $2\,050$. This challenge implies, among other things, to be able to predict the energy performance of existing or future buildings, a task which can be tackled by using thermal building models, implemented into numerical simulation computer codes.

An important example is {\em retrofit analysis}, wherein stakeholders are tasked with investigating and recommending specific refurbishment measures from vast arrays of possible options, as described in \cite{Pasanisi2008}. Retrofit analysis is necessarily based on all the available information concerning the building, such as contained for instance in the building information model (BIM) (see \cite{Eastman2011}), as well as energy consumption models implemented into computer codes. 

For instance, building energy models specifically devised for the retrofit analysis of office and school buildings are described in \cite{Rysanek2012, Tian2011}. In both works, sensitivity analyses are provided under the form of standard regression coefficient (SRC) indices (\cite{Salt00}), allowing to detect inputs which are the most influent on the predicted energy consumptions, and correction factors are applied to make model predictions more similar to historical consumption data. A more principled use of historical data involves a {\em calibration} step, wherein uncertain model parameters are tuned in order for model predictions to fit the historical data as well as possible. In \cite{Heo2012}, such a calibration is performed for normative building energy models.

{\em Design} (or {\em conception}) is another important application of building energy modeling. Contrary to retrofit analysis, no historical data are available at the conception stage, and predictions must be made for a whole range of possible designs in order to choose an optimal one. Furthermore, special care should be paid to assessing the level of confidence one can have in the predictions of such models, necessary built in the absence of historical data. In \cite{Rivalin2016}, a complete uncertainty treatment framework for thermal building modeling is presented, following the popular `non-intrusive approach' commonly used in the industrial practice \cite{Pasanisi2012a, Baudin2016}. This includes: probabilistic modeling of the different sources of uncertainty using expert opinion, sensitivity analysis, metamodeling to accelerate calculations (\cite{Sacks89}), and uncertainty propagation using either Monte-Carlo (or quasi Monte-Carlo) methods, or quadratic summation via Taylor's approximation.

Recently, these concerns for designing and maintaining buildings and energy systems to be as sustainable and green as possible, following a computer simulation based approach, have been extended to the neighborhood and the city level. That demands a systemic methodology, taking into account, in particular, interactions between energy uses, but also transportation, water and waste management (\cite{Pasanisi2016}) and gives rise to challenging problems for modellers and statisticians (\cite{Mirakyan2015}). As an example of application it is worth noting the simulation platform developed for the City of Singapore (\cite{Blin2015}).

However, all the above approaches rely on the capacity of the chosen physical model, and the ensuing computer code, to mimic `well-enough' in a certain sense the actual behavior of a real building, even though most models are based on simplified equations, which only partially reflect the building geometry and the physical properties. This is a common question in science and engineering, when complex physical systems are studied by means of computer codes, either because physical experiments are unfeasible or economically too expensive. Indeed, large differences are routinely observed between physical measurements and code predictions, raising the question of how well the code is capable of reproducing the physical system. See \cite{Koh2001} for a detailed discussion, and \cite{Damblin2015} for a recent statistical overlook on this topic. Hence, agreement between these two data sets should be checked to conclude whether or not the code can be used as a surrogate of the reality. 

This task, called \textit{validation} by \cite{Bayarri2007}, has already been tackled in the field of building energy simulation in \cite{Bontemps2015}, through graphical comparisons between measurements and corresponding code predictions for quantities of interest, such as temperature or power consumption. Although easy to use, these methods remain qualitative and do not provide the causes of a poor fitting. Many works dedicated to validation therefore propose to assess the accuracy of the code predictions in a quantitative way, and taking into account all sources of uncertainty, both {\em aleatory} and {\em epistemic}, as advocated in \cite{Roy11}, and detailed below.

When modeling thermal building behavior, aleatory uncertainty is essentially due to both weather conditions and inhabitants profile, which are naturally variable. For instance, \cite{Spitz} treats this source of uncertainty in the conception stage by conducting a local sensitivity analysis, then a global sensitivity analysis, in order to identify the factors which have the most impact on code predictions. 

Epistemic uncertainty is strongly different by nature from aleatory uncertainty; it stems from a lack of precise knowledge concerning the physical laws governing the system or process under study. One example is parametric uncertainty, {\em i.e.}, the uncertainty affecting certain model parameters, seen as physical constants (\cite{Pasanisi2012b}). In a Bayesian framework, parametric uncertainty can be encoded as a probability distribution which combines a \textit{prior} belief about the value of the uncertain parameters with the information provided by the data (\cite{Bernardo+94}). The quantification of such a parametric uncertainty is called \textit{calibration} (\cite{Cam06}).

Another source of epistemic uncertainty also arises from the lack of knowledge about the degree of adequacy between the code and the physical phenomenon, as discussed above. This can be quantified within the calibration step, together with the parametric uncertainty, following \cite{Koh2001}; however this leads to a rather complicated statistical analysis. In this work, we advocate a sequential approach where, as in \cite{Cox2001}, the presence of a discrepancy between available measures and code predictions is tested. If such test is inconclusive, in the sense that no significant discrepancy is detected, then the code is considered as {\em valid}, meaning that it can adequately be used to predict the building behavior.

Once the code has been validated, its final usage can be considered. In this study, 
we address the decision problem for an energy supplier offering a new type of contract, wherein the customer pays a fixed fee chosen in advance, based on an overall energy consumption forecast. According to Bayesian decision theory (\cite{Bernardo+94,Robert+98,French2000}), we show how to choose such a fee optimally from the point of view of the supplier, in order to balance short-terms benefits with customer loyalty.

\medskip
In Section \ref{sec:overview}, we present the case study and introduce some notations.
In Section \ref{sec:Calibration}, after assuming a statistical model between the code outputs and the power field measurements, the calibration of the thermal code is performed using a Bayesian approach. In Section \ref{sec:Validation}, the validation of the code is conducted by calculating the uncertainty affecting the average electric power delivered inside the cell over the time period. In Section \ref{sec:five}, several optimal consumption forecasts are calculated using Bayesian decision theory. We stress that, even though this study is based on real experiments, the application to optimal pricing must be seen as a fictitious exercise and it is not representative of EDF billing policies. The conclusions of the work are given in Section \ref{sec:discussion}.

\section{Overview of the case study} 

\label{sec:overview}



\subsection{Dymola computer code}

The computer code we use is a dynamic building energy model, which predicts the electric power consumption needed to heat a room to a prescribed set point temperature. The room is an experimental cell from the BESTLab laboratory, a description of which is given below. The model was built using BuildSysPro, a library of components for modeling buildings and energy systems (\cite{Plessis2014}), written in the Modelica language, and used for simulation in the Dymola software. Originally developed in \cite{Elmqvist1978}, Modelica is well adapted to model large-scale structures, the behavior of which is defined through the interaction of its components, each of them described by a limited set of equations, resulting in an overall system of (usually non-linear) equations, solved by general purpose algorithms. An important aspect of this system-based approach is its modularity, enabling to create quickly and simply large-scale models from a library of elementary building blocks. The final computer code itself is an executable built using the Dymola software.

Notationwise, let $P_t$ be the actual electric power consumption in the cell at time step $t$, for $t \in \{1, \ldots, T\}$, and let $y_t$ be the corresponding code prediction. The latter is computed given the power consumption $y_{t-1}$ at the previous time step, and a vector $\mb x_t \in \mathbb R^n$ of forcing variables (physical inputs) at time step $t$, including cell characteristics, average temperature inside the cell, and weather conditions such as wind speed, outside temperature and solar radiations, among others. In addition, this code depends on a vector of fixed parameters $\ms \theta \in \mathbb R^p.$ The dimension of $\ms\theta$ is $p=193$, including both some physical parameters such as albedo, convective factor, and many design variables such as floor surface, window width and so forth. 

Consequently, the code can be denoted by the function $g_{\ms\theta}$ such that for $t=1, \ldots, T$:
\begin{eqnarray*}
y_t &=& g_{\ms\theta}(y_{t-1}, \mb x_t).
\end{eqnarray*}
In other words, the model-defining equations are solved iteratively at each time-step, reflecting how the cell reacts dynamically to the variations of its environment. 

Equivalently, the resulting dynamic code can be described in a simpler way as a function $G_{\ms\theta}$ depending on model parameters $\ms \theta$, computing a power consumption forecast $\ms Y = (y_1, \ldots, y_T) \in \mathbb R^T$ over the whole time period, given the initial power consumption $y_0$, the complete matrix $\mb X = (\mb x_1, \ldots, \mb x_T) \in \mathbb R^{T\times n}$ of forcing variables, so that:
\begin{eqnarray}\label{eq:code_predictions}
\ms Y(\ms\theta) &=& G_{\ms\theta}(y_0,\mb X),
\end{eqnarray}
where we insist on the dependence of code predictions $\ms Y(\ms\theta)$ on the parameter $\ms\theta$, rather than on the initial condition $y_0$ and forcing variables $\mb X$, which will be held fixed throughout our study.

Note however that defining $\mb X$ as input and $\ms Y$ as output of the physical model is in a sense arbitrary, since the code essentialy solves a system of equations for which certain variables have been fixed. Hence, it would be equally possible to re-define the code as computing the temperature as a function of the electric power, as in \cite{Bontemps2013}. Our choice will be made clearer when describing the particular experimental sequence we have considered here.

\subsection{Experimental data}

\begin{figure}
\centering
\includegraphics[width=.7\textwidth]{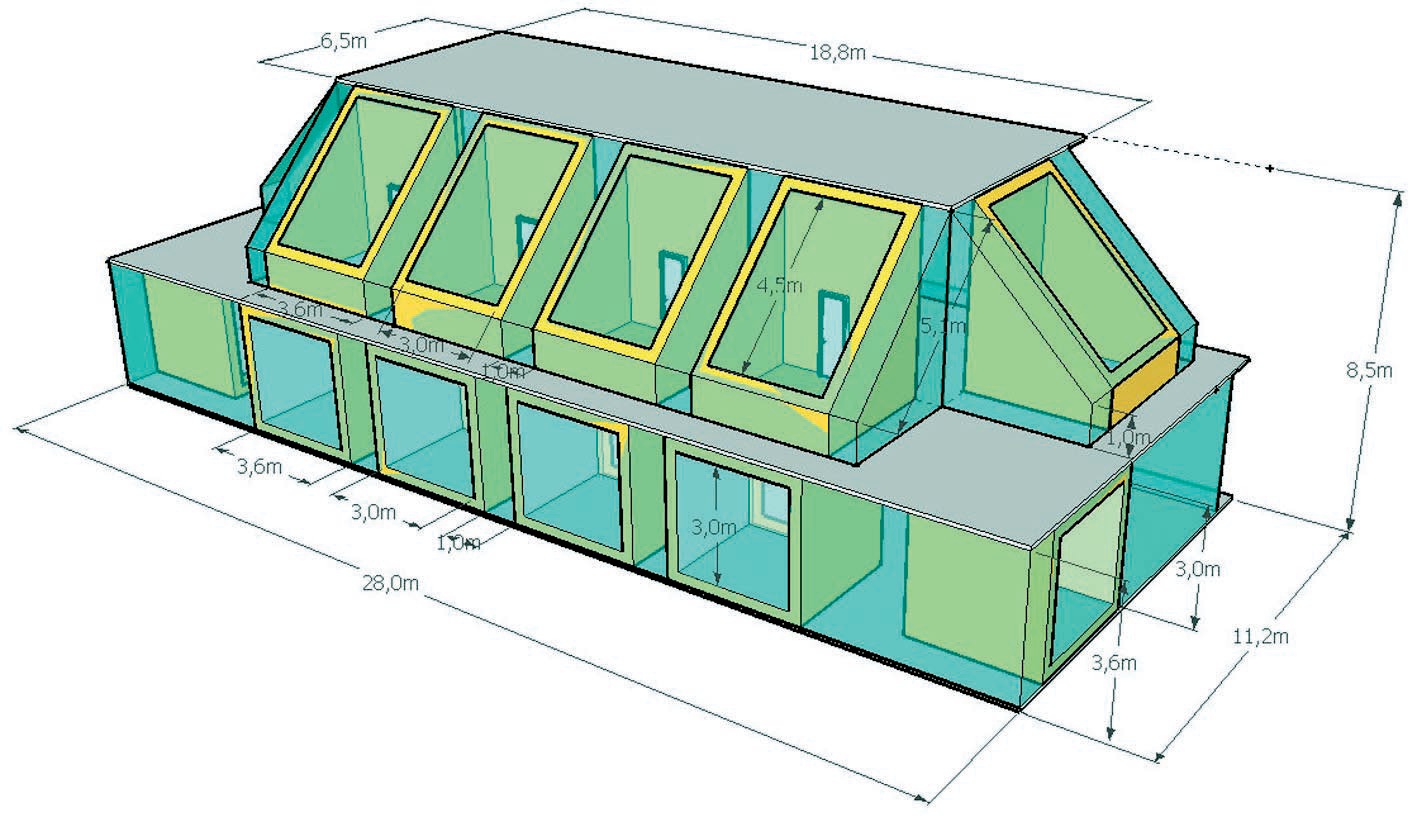}
\caption{\label{fig:bestlab_platf} Overview of the EDF BESTLAB platform.}
\end{figure}

The experimental cell is one of the twelve cells making up the BESTLab laboratory (see Figure~\ref{fig:bestlab_platf}), in the EDF research and development site of \textit{Les Renardi\`eres}, at about 75 km southeast of Paris. BESTLab was built in 2010, in the context of increasing demand for low-energy buildings. Its primary purpose is to test innovative building envelope components and integrated solar technologies. See \cite{Bontemps2013,Bontemps2015} for a detailed description of the installation. 

\begin{figure}
\centering
\includegraphics[width=0.7\textwidth]{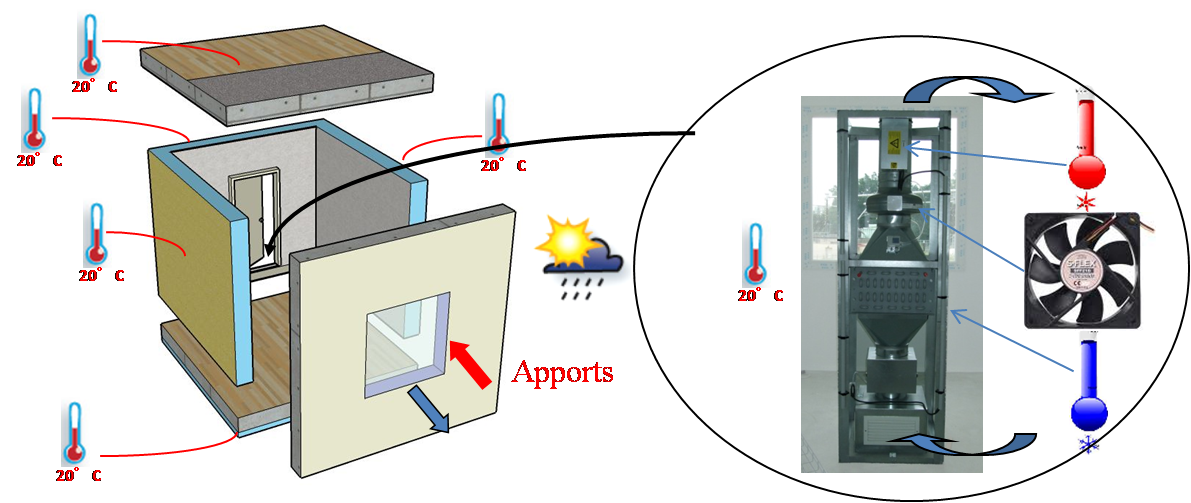}
\caption{\label{fig:cell_illu} Left: the experimental cell. Right: the HVAC system inside.}
\end{figure}

\begin{figure}
\centering
\includegraphics[width=\textwidth]{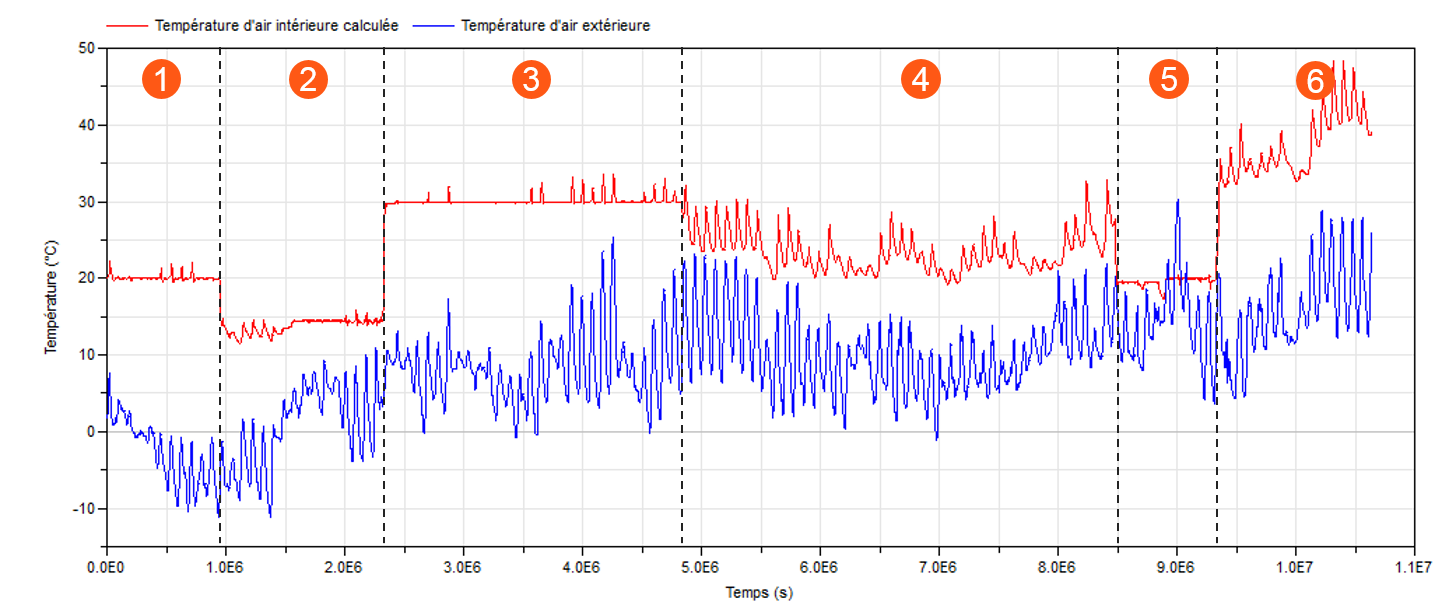}
\caption{\label{fig:full_sequence} Four month experimental sequence, comprising six periods: temperature maintained at $20^{\circ}C$ ($7$ days), $15^{\circ}C$ ($15$ days), $30^{\circ}C$ (29 days), no air conditioning (42 days), temperature maintained at $20^{\circ}C$ (7 days), heating with a constant power of $160W$ (20 days). Red: inside temperature, blue: outside temperature.}
\end{figure}

Our study focuses on one of the ground cells (see Figure~\ref{fig:cell_illu}), comprising a single, windowed, wall in contact with the outside environment. A heating, ventilation and air conditioning (HVAC) system allows to heat or cool the cell at will, thus allowing to test different types of scenarios. Figure~\ref{fig:full_sequence} shows a typical experimental sequence, wherein the cell is alternatively maintained at certain target temperatures, heated with a fixed power supply, or left unconditioned, so that the temperature evolves freely. 

In our case, we will focus on the first $7$ days period of this sequence, where the cell is maintained at $20^{\circ}C$. The main reason for doing so is that this experimental condition resembles most the usual state of residential housing. Furthermore, the code may reflect more accurately certain experimental conditions then others, hence it makes more sense to validate it separately for each experimental condition, before considering a global validation. 

Moreover, heterogeneity in the experimental sequences raises the issue of defining which are the inputs and outputs of the computer code, as discussed in the previous section. Indeed, when maintaining the cell at a fixed temperature, it makes sense to consider the latter as an input, and the power consumption as the output of interest. In contrast, when the cell is heated with a fixed power supply, or when the HVAC system is turned of, it would make more sense from a physical point of view to consider the power consumption as an input (which may be equal to zero), and the resulting temperature as an output.

\begin{figure}
\centering
\includegraphics[width=0.5\textwidth]{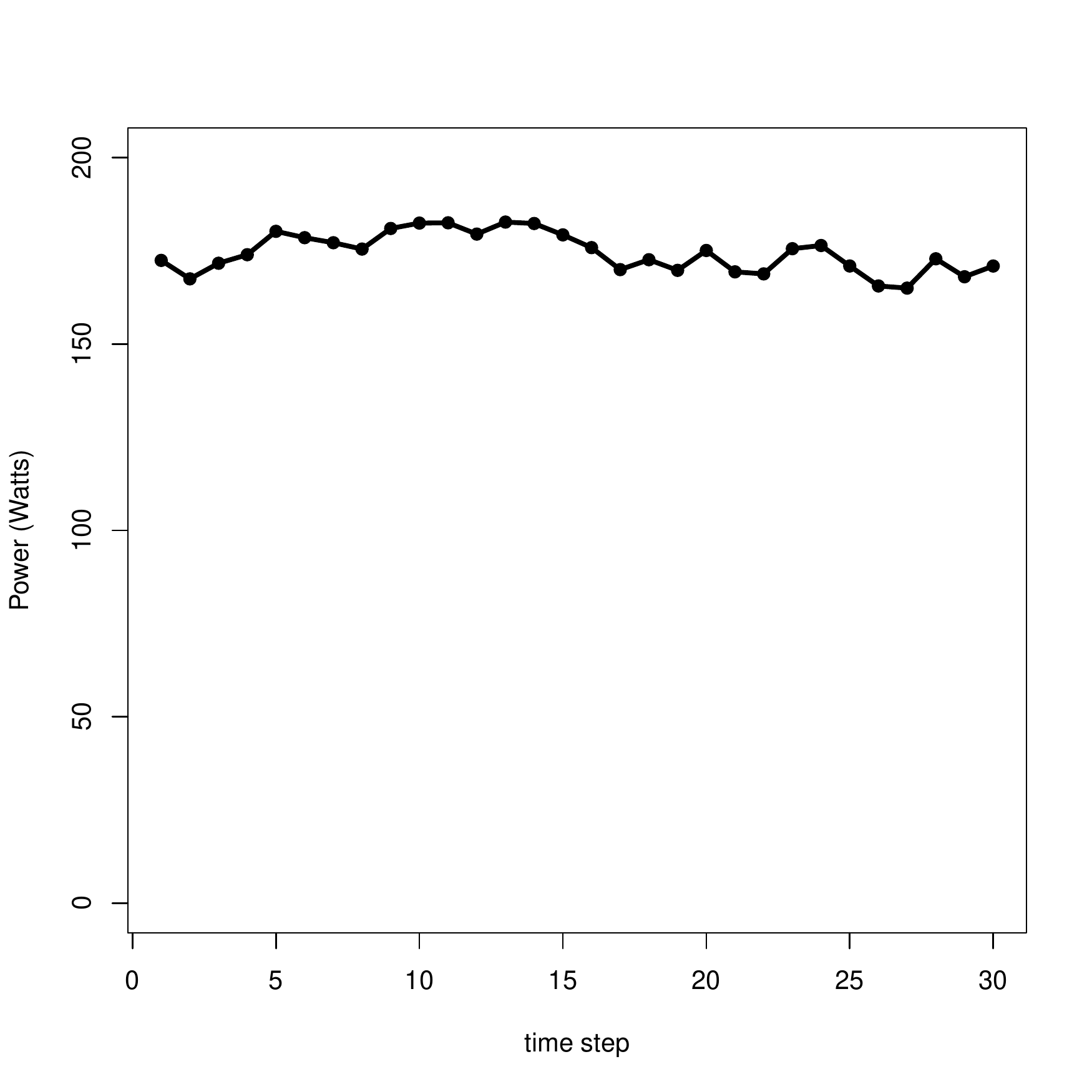}
\caption{\label{fig:fit_field_simu_moy} The $30$ averaged power measurements. The time step is approximately $5$ hours and $30$ minutes.}
\end{figure}

Finally, because our final use of the code is to predict the total power consumption over a given period, the measured as well as computed powers are averaged, resulting in $30$ time steps over $7$ days, rather than the $2016$ five-minute time-steps initially considered. This allows to considerably reduce the computational burden of the analysis, while also reducing the variance, both in the experimental and the computer code outputs. Figure \ref{fig:fit_field_simu_moy} shows the resulting smoother data we use for code validation.

\subsection{Observation model}

We assume in the following that no aleatory uncertainty taints the inputs $\mb X$ of our code, since they have been observed throughout the experiment and are thus considered as fixed quantities (covariables). On the other hand, certain model parameters are considered uncertain, including the thermal bridge factor, the albedo and the convective factor associated with the HVAC system. 

The power field measurements are denoted by the vector $\mb Z = (z_1, \ldots, z_T)$. As discussed above, computations are performed based on the averaged rather than complete power measurements, resulting in about four measures per day instead of one per $5$ minutes, yielding $30$ data over the $7$ days. Owing to measurement errors, the power measurements $z_t$ are not exactly equal to the true power consumptions $P_t$. Modeling these errors as a white noise, we have:
\begin{eqnarray}
\label{eq:real}
z_t &=& P_t + \epsilon_t\\
\epsilon_t &\underset{i.i.d.}\thicksim& \mathcal{N}(0,\lambda^2),\nonumber
\end{eqnarray}
where $\epsilon_t$ is a zero mean Gaussian random variable, with variance $\lambda^2$ which can also account for residual variability (\cite{Koh2001}) as well as measurement errors.

\subsection{Sensitivity Analysis}

Given that the dimension of $\ms\theta$ is very large ($p=193$), direct calibration of the thermal code is not feasible. 
Hence, a preliminary screening step is necessary in order to remove parameters that have a negligible influence on power predictions (\cite{Salt00}). To this end, we have used the results from \cite{Bontemps2013}, which we describe now.

In a first step, a local sensitivity analysis was conducted, using a one-at-a-time (OAT) approach, consisting in computing the deviations in code outputs at each time step when a single parameter was changed by $5$ percent around its nominal value. The corresponding sensitivity indices were defined as the ratios of output over input deviations~:
\begin{eqnarray*}
\ms S_i &=& \frac{\ms Y(\ms\theta^\ast+h_i \ms e_i)-\ms Y(\ms\theta^\ast-h_i \ms e_i)}{2 h_i} \approx \frac{\partial \ms Y(\ms\theta)}{\partial \theta_i}|_{\ms\theta=\ms\theta^\ast},
\end{eqnarray*}
where $\ms\theta^\ast$ is the nominal value for code parameters, elicited from expert opinion, and $\ms e_i = (0,\ldots,1,\ldots,0)$ is the $i$-th vector of the canonical orthonormal basis for $\mathbb R^p$. $\ms S_i$ is approximately equal to the code's partial derivative with respect to the $i$-th parameter, computed at each time-step (keeping in mind that $\ms Y(\ms\theta)$ is a vector of size $T$). Both the mean $S_{i,m}$ and standard deviation $S_{i,std}$ where computed over time, resulting in a hybrid index $S_{i,d} = \sqrt{S_{i,m}^2 + S_{i,std}^2}$, allowing to retain parameters with a low mean, but highly variable influence on the power predictions. By thresholding this indicator, the number of parameters was downsized from $p=193$ to $p=13$.

In a second step, a global sensitivity analysis of the remaining parameters was performed, which consisted in computing their respective time-varying Sobol indices, using a Monte-Carlo approach coupled with a polynomial chaos expansion response surface (for further details, see \cite{Bontemps2013}). This second analysis showed that the sum of first-order indices were always higher then $97\%$, suggesting an absence of interaction between parameters at all time. Furthermore, the maximum values over time of the first order Sobol indices were used to identify three parameters, with maximum values over $25\%$, the remaining parameters having maximum Sobol indices below or equal to $7\%$.

Finally, the three scalar parameters having the greatest impact on the output were found to be:
\begin{itemize}
\item $\theta_1\in[0,1]$ which is the albedo factor,
\item $\theta_2>0$ which encodes the effect of the thermal bridges, 
\item $\theta_3>0$ which is the convective factor of the HVAC system.
\end{itemize}



\section{Calibration}

\label{sec:Calibration}


As discussed above, the thermal code depends on a vector $\ms \theta$ of physical parameters, typically set to an unchanged value before running the code, for instance a nominal value $\ms \theta^\ast$ set by experts. However, $\ms \theta$ is often uncertain, typically because it is non-measurable in the field. 
Code calibration consists in reducing this (epistemic) parametric uncertainty, by identifying parameter values for which code outputs are as close as possible to physical measures.

In calibration, it is important to keep in mind that the code might be an imperfect representation of the physical system. Hence, it should be considered as a more or less accurate mathematical approximation of the thermal behaviour inside the experimental cell. This second source of epistemic uncertainty is called {\em code uncertainty}, and is dealt with in \cite{Koh2001} by adding a {\em discrepancy} term in the statistical model used for calibration. We advocate here another approach, based on a {\em post-hoc} statistical test to detect the presence of such a discrepancy between measures and code predictions (see Section~\ref{sec:Validation}).

Adopting a Bayesian perspective, calibration requires the following ingredients:
\begin{itemize}
\item a statistical model which links the available field measurements $\mb Z$ with the code outputs. This equation provides a likelihood function $\mathcal{L}(\mb Z|\ms\theta,\mb\psi)$, where $\mb\psi$ is a vector of nuisance parameters attached to the model, specifying for instance the error structure between code outputs and field measurements.
\item A prior density $\pi(\ms \theta )$ encoding the uncertainty as a \textit{prior} belief in favor of some values of $\ms\theta$, which are more probable than others, based on expert opinion. If no such prior information is available, a uniform \textit{prior} distribution can be adopted.
Similarly to $\ms\theta$, $\mb\psi$ is endowed with a prior density $\pi(\ms \psi)$, which we choose independent from $\pi(\ms \theta )$, meaning that we form {\em a priori} independent opinions about the plausible values of both parameters.
\end{itemize}

The prior uncertainty affecting both code  $\ms\theta$ and nuisance $\mb\psi$ parameters is then updated according to Bayes' theorem:
\begin{align}
\nonumber
\pi(\ms\theta,\mb\psi|\mb Z)=&\frac{\mathcal{L}(\mb Z|\ms\theta,\mb\psi)\pi(\ms\theta)\pi(\mb\psi)}{\int_{\ms\theta,\mb\psi} \mathcal{L}(\mb Z|\ms\theta,\mb\psi)\pi(\ms\theta)\pi(\mb\psi)\dd\ms\theta\dd\mb\psi}\\
\propto &\,\,\,\,\,\, \mathcal{L}(\mb Z|\ms\theta,\mb\psi)\pi(\ms\theta)\pi(\mb\psi).
\label{Bayesformula}
\end{align}

We now detail the choice of the statistical model and priors.

\paragraph{Statistical model.}

Assuming the model discrepancy is negligible, {\em i.e.} the code is a faithful representation of the cell's behavior, we have:
\begin{equation}
\label{eq:unbiased}
\exists \ms\theta_0\in\mathcal{T}\,\,\,s.t.\,\,\,\ms Y(\ms\theta_0)=\ms P,
\end{equation}
with $\ms P = (P_1, \ldots, P_T)$ the sequence of `true' power consumptions inside the cell, which remain unkown. Then, (\ref{eq:real}) implies that:
\begin{equation}
\label{eq:modelstat}
\mb Z = \ms Y(\ms\theta_0) + \ms\epsilon,
\end{equation}
where $\ms \epsilon = (\epsilon_1, \ldots, \epsilon_T)$ is defined by (\ref{eq:real}). Since $\ms\epsilon_t$ is a Gaussian white noise, the likelihood 
is given by:
\begin{equation*}
\mathcal{L}(\mb Z|\ms\theta,\lambda^2)=\frac{1}{(\sqrt{2\pi}\lambda)^{n}}\exp{\Big[-\frac{1}{2\lambda^2}SS(\ms\theta)\Big]},
\end{equation*}
where 
\begin{equation*}
SS(\ms\theta)=||\mb Z- \ms Y(\ms \theta)||^{2}
\end{equation*}
is the sum of squares of the residuals between the power field measurements and the code outputs. 

\paragraph{Prior densities.}

The only nuisance parameter appearing in our case is the variance $\lambda^2$ of observation errors, as defined by~(\ref{eq:real}), hence $\psi := \lambda^2$.

Based on the minimal available information, we defined $\pi(\ms\theta)$ as a product of independent uniform distributions, for which bounds were chosen by thermal modeling experts in order for the parameter values to remain physically plausible. For instance, the albedo $\theta_1$ being a reflection coefficient, it is necessarily between $0$ and $1$. Furthermore, a Jeffrey non-informative \textit{prior} is specified on the variance $\lambda^2$, yielding:

\begin{equation*}
\pi(\ms\theta)=\pi(\theta_1)\pi(\theta_2)\pi(\theta_3)
\end{equation*}
where
$$\pi(\theta_1)=\frac{\mathbf{1}_{[0,1]}(\theta_1)}{1}
\,\,\,\,\pi(\theta_2)=\frac{\mathbf{1}_{[0,100]}(\theta_2)}{100}
\,\,\,\,\pi(\theta_3)=\frac{\mathbf{1}_{[0,100]}(\theta_3)}{100},
$$
and
$$
\,\,\,\,\pi(\lambda^2)\propto\frac{1}{\lambda^2}.$$

\paragraph{Posterior distribution.}

\begin{figure}
\centering
\includegraphics[width=0.4\textwidth]{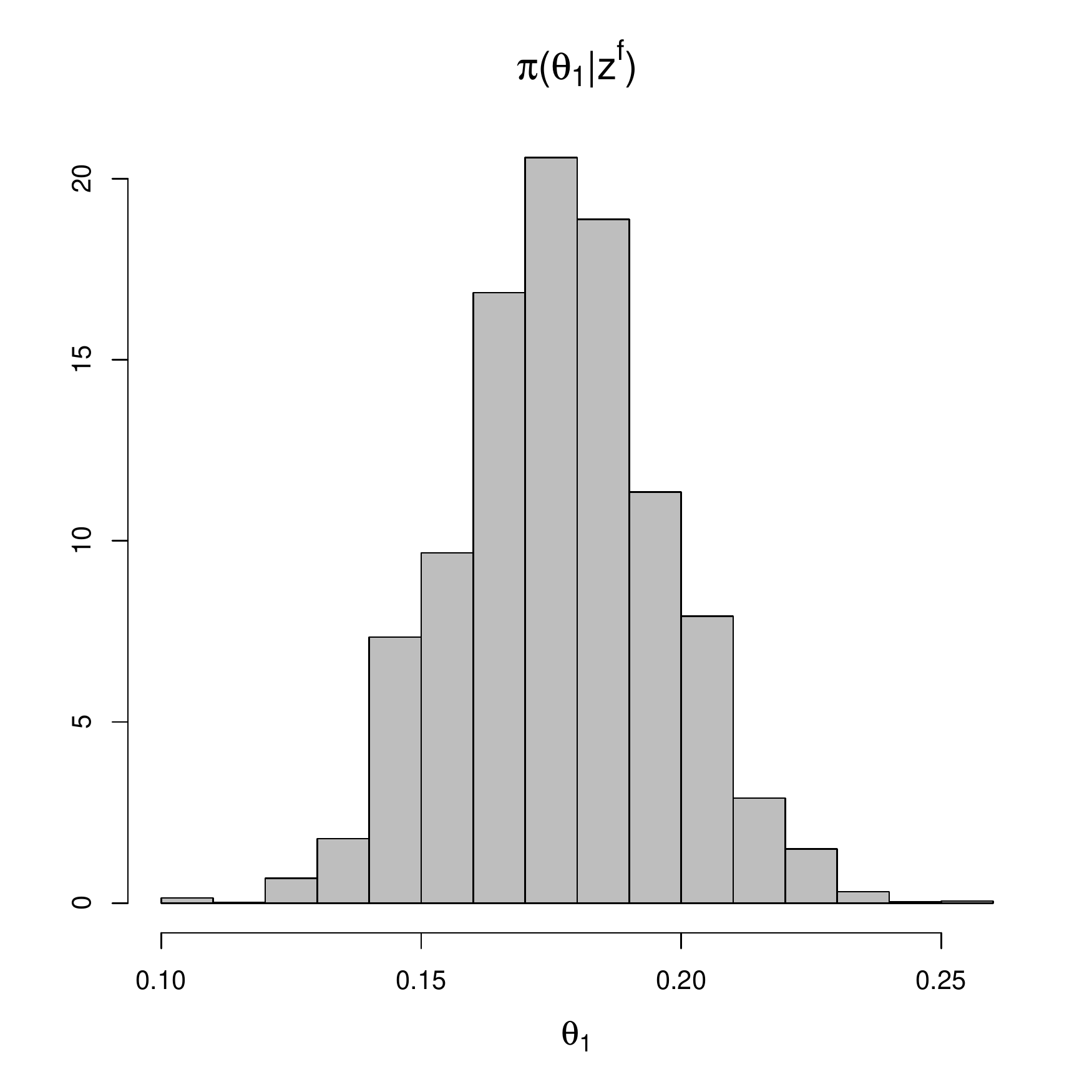}
\includegraphics[width=0.4\textwidth]{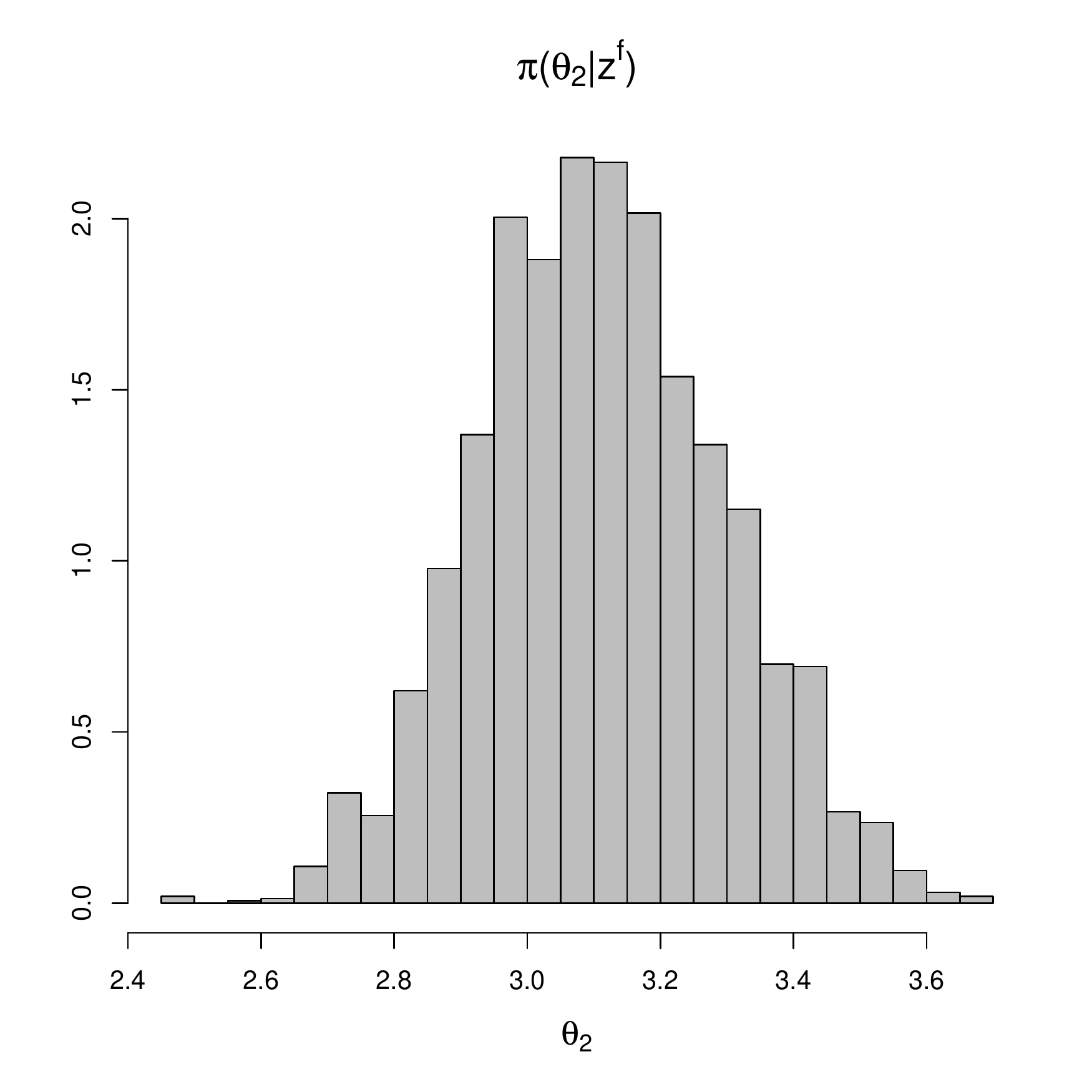}\\
\includegraphics[width=0.4\textwidth]{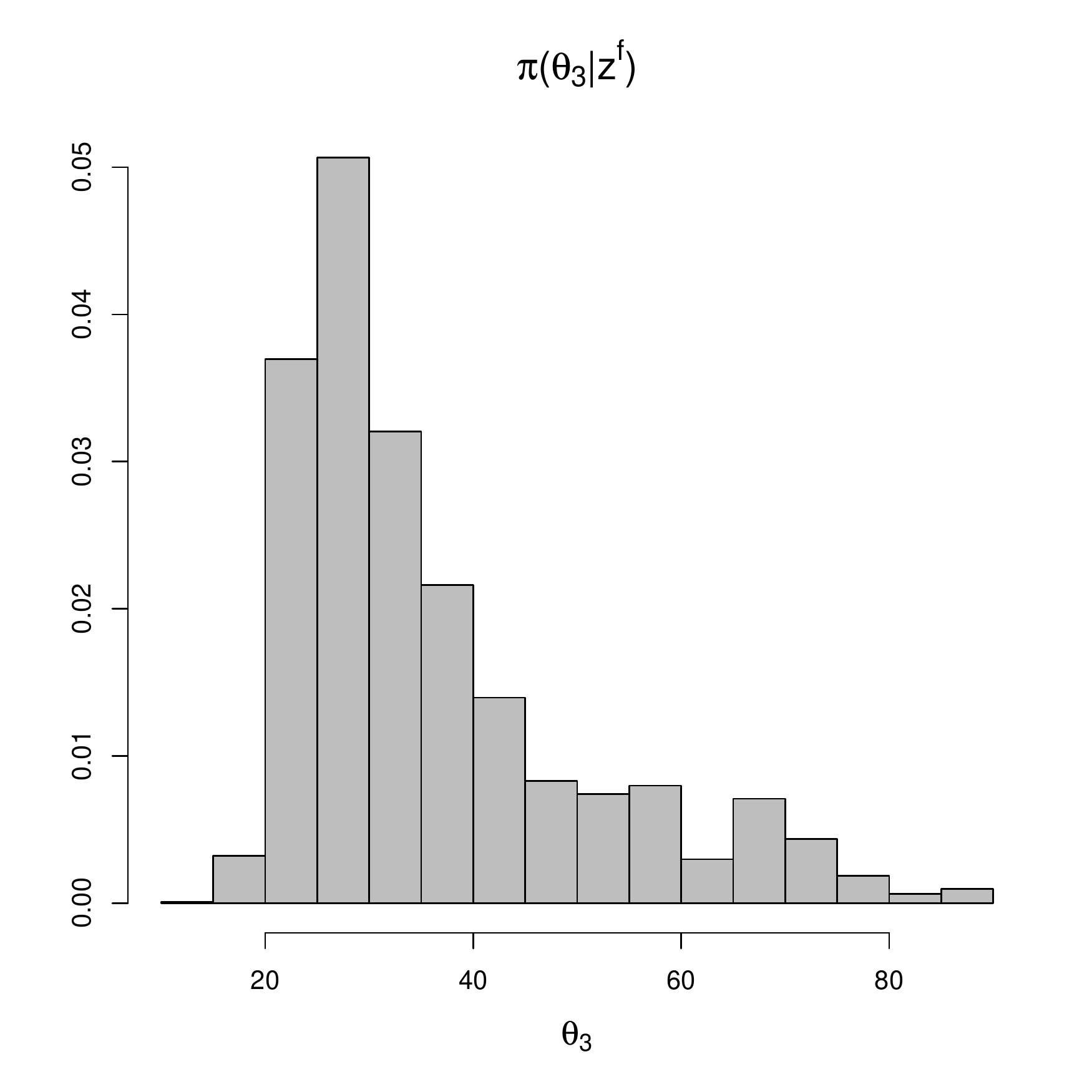}
\includegraphics[width=0.4\textwidth]{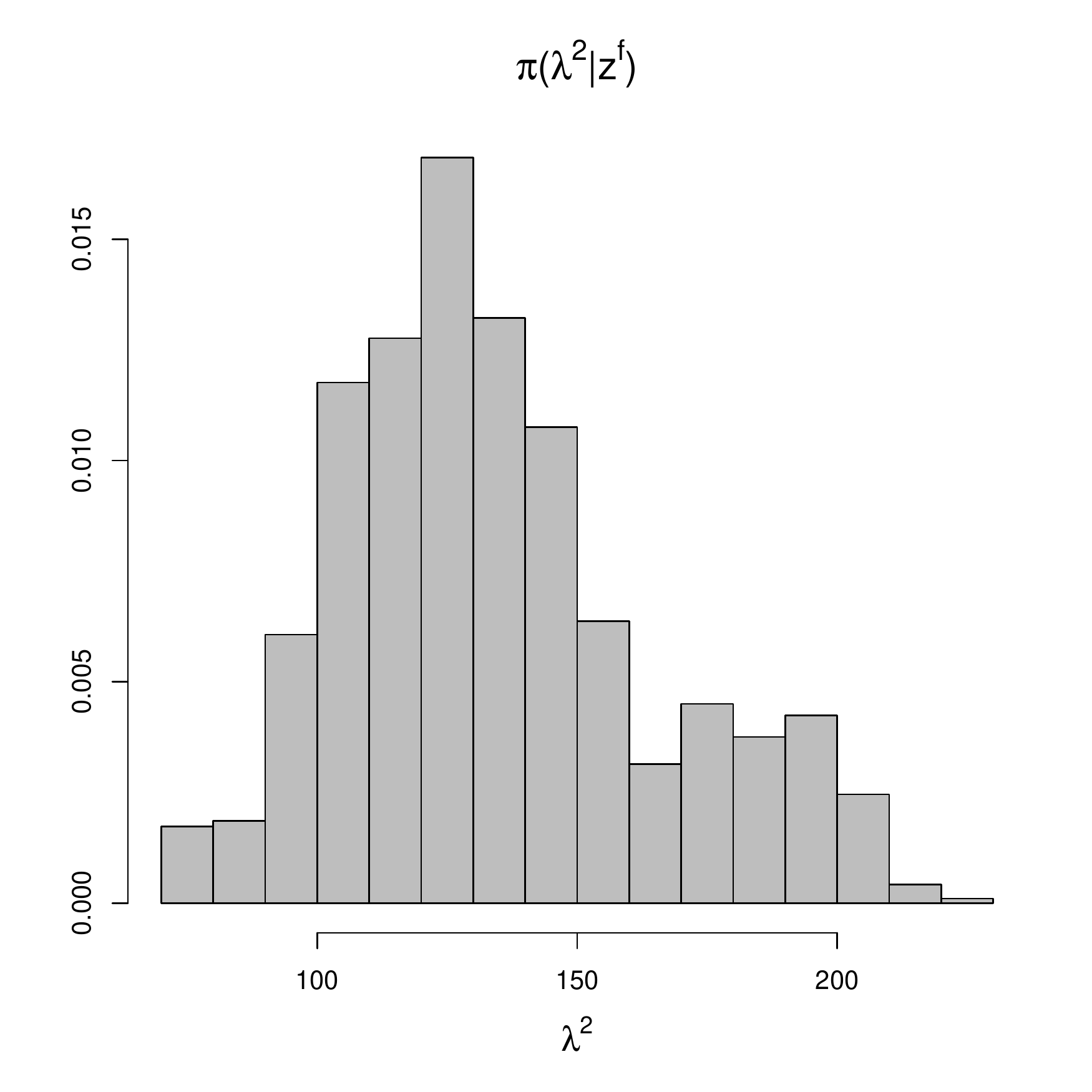}
\caption{\label{fig:post} From left to right and top to bottom : marginal posterior densities of $\,\theta_1, \,\theta_2,\,\theta_3$ and $\,\lambda^2$.}
\end{figure}

From Bayes' formula (\ref{Bayesformula}),
\begin{equation}
\label{eq:Bayes_formula_thermal_code}
\pi(\ms\theta,\lambda^2|\mb Z)\propto \mathcal{L}(\mb Z|\ms\theta,\lambda^2)\pi(\ms\theta,\lambda^2).
\end{equation}
As $\ms Y\!:\!\ms \theta \mapsto \ms Y (\ms \theta)$ is non-linear, the \textit{posterior} distribution (\ref{eq:Bayes_formula_thermal_code}) has no closed form. It is sampled using a Metropolis-Hastings algorithm (\cite{Robert+98}). In practice, calculations were carried out using the algorithm implemented in the Open-TURNS software platform for uncertainty treatment in numerical simulation (\cite{Baudin2016}), consisting of componentwise random-walk steps with adaptive lengths, tuned to maintain the acceptance rate within reasonable bounds.

Figure~\ref{fig:post} shows the densities of the marginal \textit{posterior} distributions $\pi(\ms{\theta_i}|\mb Z)$ and $\pi(\lambda^2|\mb Z)$. These show that the $30$ available data allowed to significantly reduce the parametric uncertainty, yielding posterior densities with masses concentrated in realistic regions of the parameter space. For instance, the albedo is seen to be with high probability between $0.15$ and $0.20$, values usually associated to bare soil and green grass, which happens to be precisely the environment of the BESTLAB platform.

\section{Validation}
\label{sec:Validation}



Validation is the second task of the V$\&$V (Verification and Validation) framework aiming at quantifying the accuracy of code predictions (\cite{AIAA}). Verification consists in checking that all bugs inside the source code have been removed and, when the code is constructed as a finite element solution, that the discretization errors are small enough (\cite{Roache1998}). In this paper, we are not concerned with verification although this task should be addressed before any validation study. 

The goal of validation is to ensure that the mathematical representation that underlies the code is an acceptable representation of the physical system. 
A naive approach, yet still very popular in day to day industrial practice, consists in  visually comparing code outputs and field measurements, and then decide if the difference is small enough. Over the past decade, much effort has been put into quantifying more rigorously this difference and the uncertainty affecting it. For instance, \cite{Roy11} introduced several validation metrics, based on statistical tests taking into account that the computer code may be subject to different types of error, such as: intrinsic input variability (aleatory uncertainty), and parametric (epistemic) uncertainty. 

\cite{Bayarri2007} proposed a Bayesian validation framework in which the discrepancy between the code and the field measures is modeled by a random Gaussian process, following the seminal idea of \cite{Koh2001}. In our study, the aleatory uncertainty is negligible because $\mb{x}_t$ is measured precisely at each time step $t$ over the time period and the measurement error attached to it is included in the noise term $\epsilon_t$ (\ref{eq:real}). On the other hand, the power predictions are affected by both the value of the code parameters and the adequacy of the code itself to the thermal system. The validation stage is thus closely dependent on the results of the calibration stage.  In \cite{Bayarri2007}, the epistemic uncertainty is quantified during the calibration stage, and then propagated to the code output. We will follow the same idea, except for the fact that we do not use a Gaussian process modeling a possible code discrepancy. We rely instead on a statistical test of the presence of such a discrepancy.

Last but not least, an important point is that a validation study should be carried out keeping in mind the intended use of the code (\cite{AIAA}). Consequently, code predictions should be judged sufficiently accurate according to the estimation of a quantity of interest $\phi$ reflecting its intended use.

In this study, we merely aim at assessing the uncertainty affecting the average power delivered inside the cell over the chosen time period. Hence,
\begin{eqnarray}
\label{eq:mean}
\phi(\ms P) &:=& \bar {\ms P} = \frac{1}{T}\sum_{t=1}^{T} P_t.
\end{eqnarray}
It follows that the uncertainty affecting $\bar{\ms P}$ derives from the uncertainty affecting $P_t$ at each time step over the period.

\begin{figure}
\centering
\includegraphics[scale=0.50]{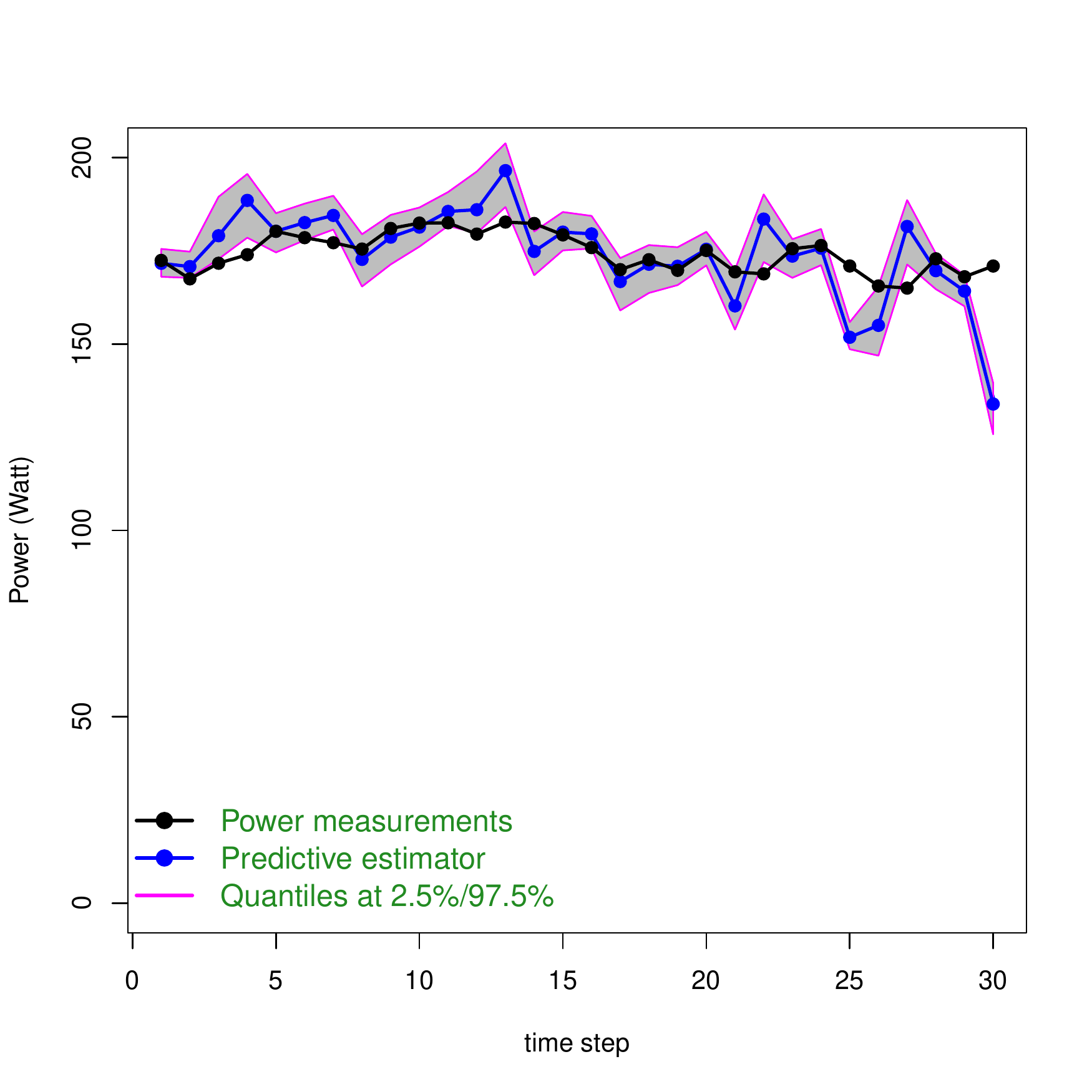}
\caption{\label{fig:pred_puiss} Measures versus calibrated code predictions of the electric power delivered inside the cell at each time step of the period.}
\end{figure}

\begin{figure}
\centering
\includegraphics[scale=0.50]{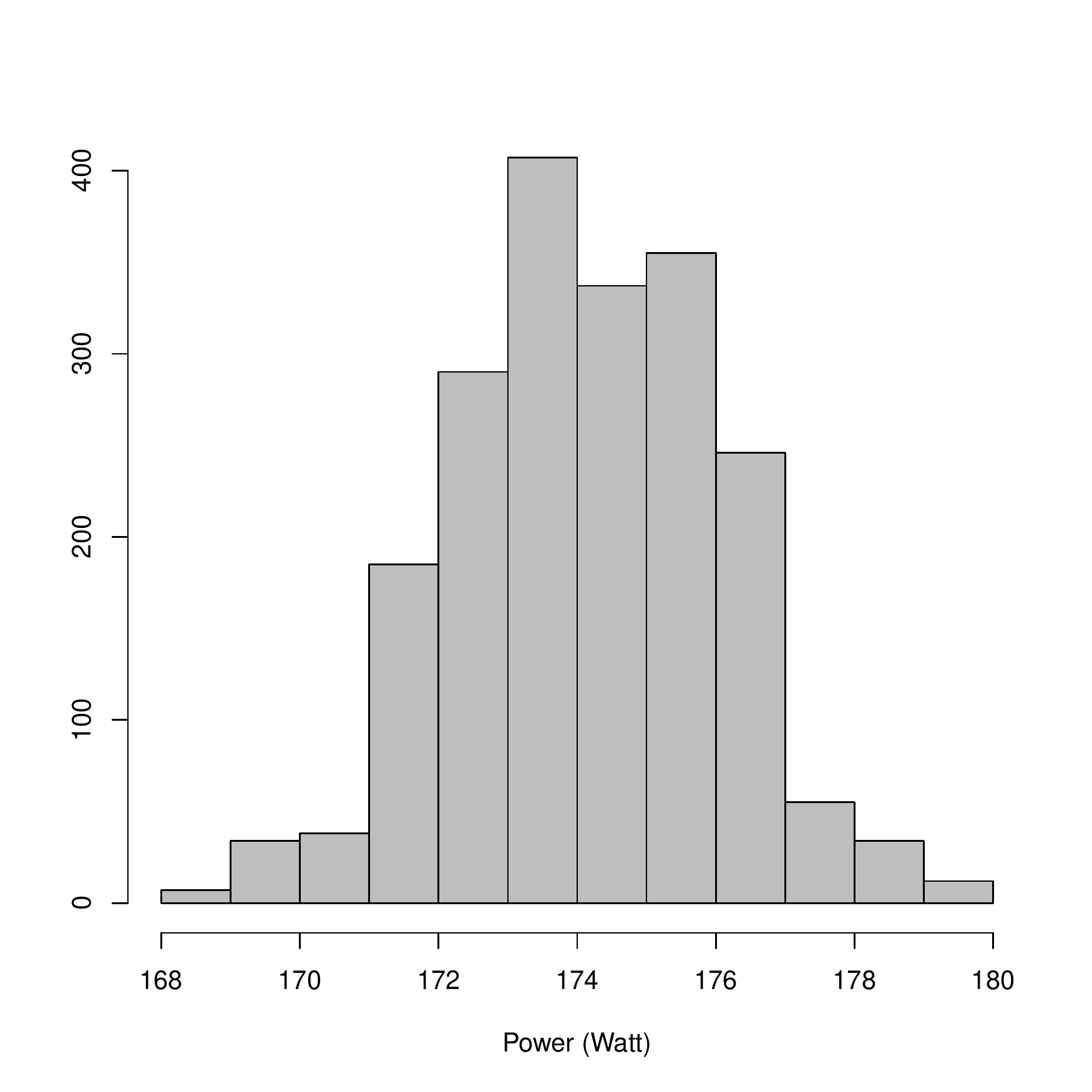}
\caption{\label{fig:pred_puiss_moy}Probability distribution of the averaged electric power delivered inside the cell over the time period.}
\end{figure}

As detailed in Section~\ref{sec:overview}, validation of the thermal code consists in assessing the uncertainty affecting the average power $\bar{\ms P}$ which is delivered inside the cell over the time period. This process requires the followings steps:
\begin{enumerate}
\item Generate a sample $(\ms\theta_1,\cdots,\ms\theta_M)$ from the posterior distribution $\pi(\ms\theta|\mb Z)$, as described in Section \ref{sec:Calibration};
\item Run the code over the $M$ samples $(\ms\theta_1, \ldots, \ms\theta_M)$\footnote{Actually, these runs are necessarily performed during the calibration step. A good practice is therefore to store all the computer code evaluations done during calibration, to avoid having to do them all over again for the validation.}. This leads to a sample $(\ms Y (\ms\theta_1), \ldots, \ms Y (\ms\theta_M))$ from the the posterior distribution $\pi(\ms Y(\ms\theta)|\mb Z)$ of the electric power over the time-period. A point estimate can then be derived, such as the posterior mean:
\begin{equation*}
\mathbb E[\ms Y(\ms \theta) | \mb Z] = \int \ms Y (\ms\theta) \pi(\ms\theta|\mb Z)\dd\ms\theta,
\end{equation*}
which we approximate here by the empirical mean $\frac{1}{M} \sum_{m=1}^M \ms Y (\ms\theta_m)$, as illustrated by the blue line in Figure~\ref{fig:pred_puiss};
\item Estimate the posterior distribution of $\pi(\bar{\ms P}|\mb Z)$ by $\pi(\bar{\ms Y}(\ms \theta)|\mb Z)$, following (\ref{eq:unbiased}), a sample of which is given by $(\bar{\ms Y} (\ms\theta_1), \ldots, \bar{\ms Y} (\ms\theta_M))$, where the upper bar denotes the mean over the time period, as defined by (\ref{eq:mean}).
\end{enumerate}
Then, practitioners should decide whether or not the uncertainty on $\bar{\ms P}$ is not too large in view of its intended use, for instance by calculating the quantiles at $95\%$ of $\pi(\bar{\ms P})$ (see Figure \ref{fig:pred_puiss_moy}, where we see that the average power consumption is most likely between $170\,W$ and $178\,W$ ). 

\subsection{Statistical testing}

To summarize, the above validation procedure consists of a Bayesian calibration step, and two consecutive uncertainty propagation steps, which depend crucially on the calibration result, that is, the probability distribution $\pi(\ms\theta|\mb Z)$, which depends itself on the statistical model chosen for calibration. In the litterature dedicated to code calibration, as we have discussed previously, an alternative to the assumption of unbiasedness of the code with respect to reality (\ref{eq:unbiased}) is given by:
\begin{eqnarray}
\label{eq:biased}
\ms Y(\ms\theta_0) &=& \ms P + \ms b,
\end{eqnarray}
where $\ms\theta_0$ is often defined as a `best-fitting' value (\cite{Koh2001}) with respect to the available data. This amounts to saying that there exists no value of $\ms\theta$ making a perfect agreement between $\ms P$ and $\ms Y(\ms\theta)$ . The residual gap $\ms b$ is usually called {\em code discrepancy}. Joint estimation of $\ms\theta$ and $\ms b$ is performed in \cite{Bayarri2007}, despite several conceptual as well as practical difficulties:
\begin{itemize}
\item Assumption (\ref{eq:biased}) leads to a statistical model wherein only $\ms Y(\ms\theta) + \ms b$ can be estimated; this results in a confoundment between $\ms\theta$ and $\ms b$ which is resolved in the literature by adopting a Gaussian process (GP) prior distribution on $\ms b$. This means that the estimated value of $\ms\theta$ depends entirely on the prior chosen for $\ms b$, irrespective of the number of available data;
\item The definition of the `real value' $\ms \theta_0$ for the parameter becomes problematic as soon as the code itself is no more considered an exact depiction of reality;
\item Adding the GP term makes model estimation as well as prediction more complex from a purely technical point of view.
\end{itemize}
All these difficulties may explain why this approach has not yet been widely adopted by the engineering community. As a consequence, we choose here to follow another path, relying on a statistical test of the null hypothesis $\mathcal H_0 : \ms b = \ms 0$, that is, that our unbiased statistical model (\ref{eq:modelstat}) is an accurate enough description of reality, vs. $\mathcal H_1 : \ms b \neq \ms 0$, meaning that their is a systematic bias in the model predictions, according to (\ref{eq:biased}).

Let us consider as a goodness-of-fit test statistic the $\mathcal{X}^{2}$ discrepancy (\cite{Gelman1996}):
\begin{equation}\label{eq:discrepancy}
\mathcal  X^2(\mb Z,\ms\theta,\lambda^2)=\frac{||\mb Z - \ms Y(\ms\theta)||^2}{\lambda^2}.
\end{equation}
The classical testing procedure then consists in computing the following $p$-value:
\begin{equation*}
p_{val}(\ms\theta, \lambda^2, \mb Z)=\pp[\mathcal X^2(\mb Z_{rep},\ms\theta,\lambda^2)>\mathcal X^2(\mb Z,\ms\theta,\lambda^2) | \ms\theta, \lambda^2, \mb Z, \mathcal H_0],
\end{equation*}
where $\mb Z_{rep}$ is a vector of replicated data, simulated under the null hypothesis, given here by our statistical model (\ref{eq:modelstat}). Given a maximum acceptable type I (false positive) error rate $\alpha$ (typically $5\%$), $\mathcal H_0$ is then rejected for values of the $p$-value smaller then $\alpha$.
If model parameters $\ms\theta$ and $\lambda^2$ {\em were perfectly known}, the null distribution would be chi-square, with $T$ degrees of freedom (df).

The usual frequentist way of dealing with the uncertainty tainting $\ms\theta$ and $\lambda^2$ is to estimate them from the data $\mb Z$ through, say, a maximum likelihood procedure. The null distribution is then approached, either by bootstrap techniques (\cite{Efron1981}) or asymptotically, leading here to the chi-square distribution with $T-(p+1)$ degrees of freedom (\cite{Vandervaart2000}).

\begin{figure}
\centering
\includegraphics[scale=0.40]{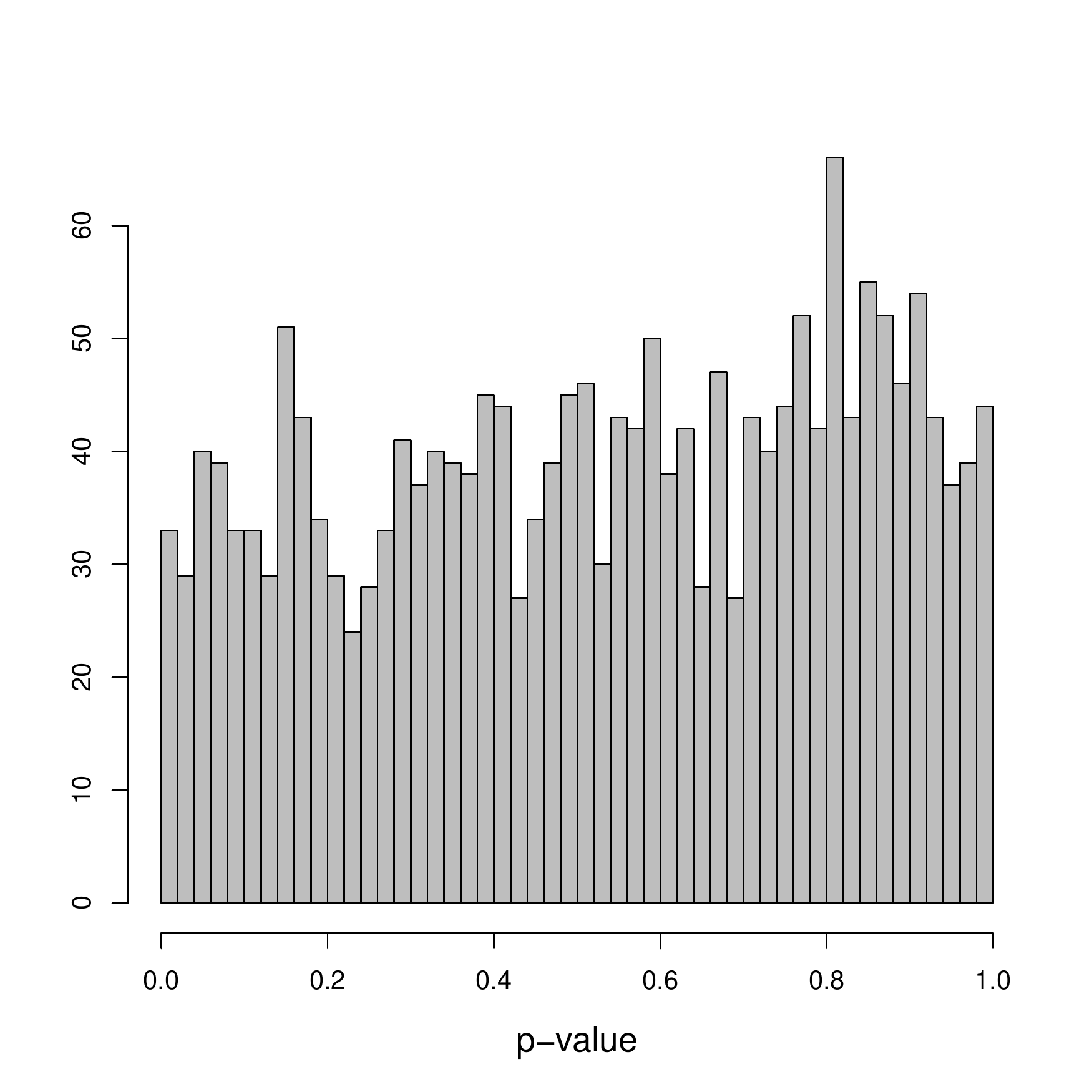}
\includegraphics[scale=0.40]{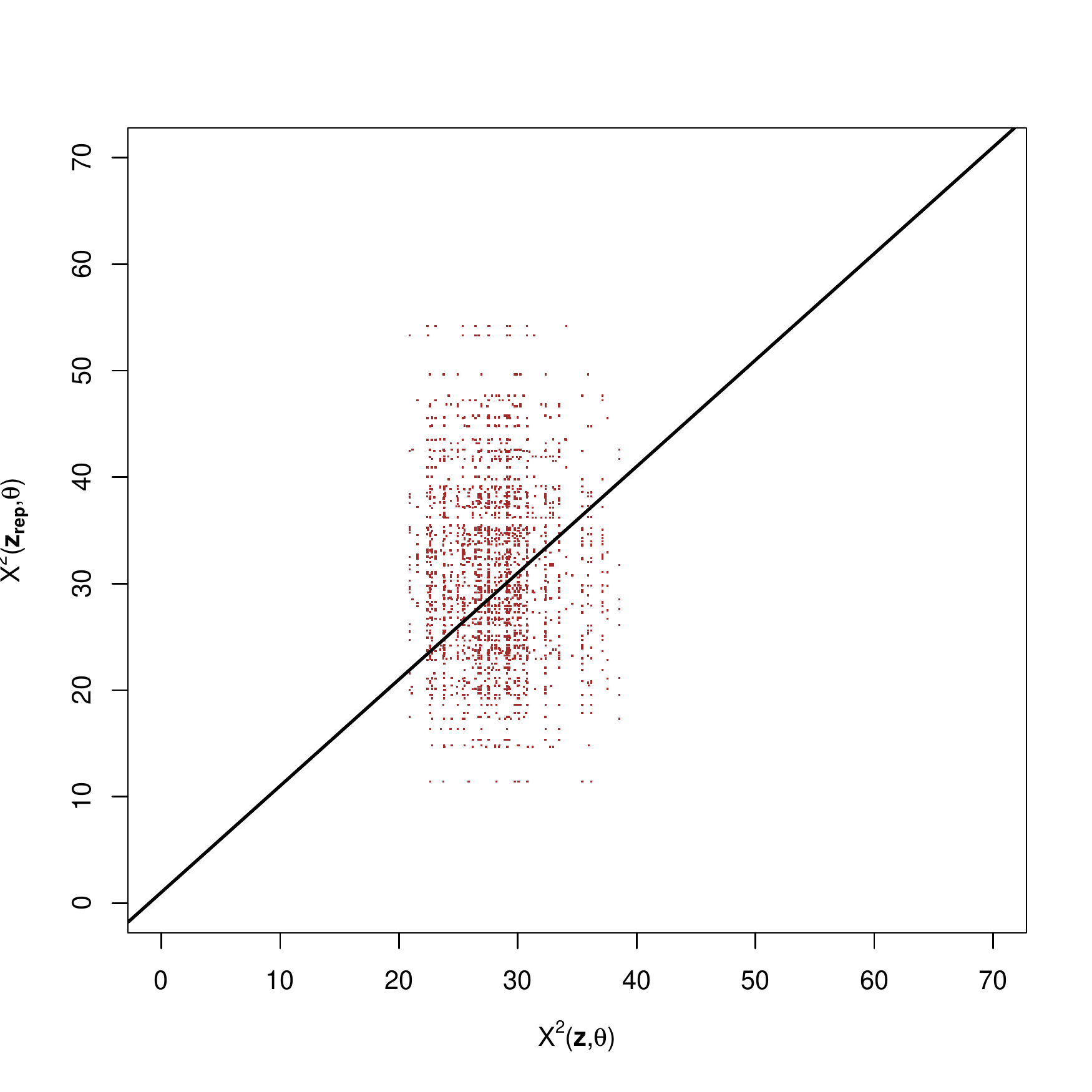}
\caption{\label{fig:scatt} Left: Histogram of the posterior sample of $p_{val}(\ms\theta, \lambda^2, \mb Z)$: the posterior $p$-value is estimated by the average over this sample. Right: Scatterplot of predictive vs. realized $\mathcal{X}^{2}$ discrepancies: the posterior $p$-value is estimated by the proportion of points above the black line.}
\end{figure}

Since we have adopted a Bayesian perspective, our way to deal with parametric uncertainty is to compute the \textit{posterior predictive} $p$-value, as introduced by \cite{Gelman1996}, which takes into account the posterior distribution of $\ms\theta$ and $\lambda^2$, following:
\begin{equation}\label{eq:post_p_value}
p_B(\mb Z)=\pp[\mathcal X^2(\mb Z_{rep},\ms\theta,\lambda^2)>\mathcal X^2(\mb Z,\ms\theta,\lambda^2)|\mb Z, \mathcal H_0]
\end{equation}
where $\mb Z_{rep}$ is a vector of replicated data, which is now simulated from the predictive density derived from (\ref{eq:modelstat}). The probability in (\ref{eq:post_p_value}) is therefore computed with respect to the joint \textit{posterior} density of $(\mb Z_{rep},\ms\theta,\lambda^2)$:
\begin{equation*}
\pi(\mb Z_{rep},\ms\theta,\lambda^2|\mb Z)=\mathcal L(\mb Z_{rep}|\ms\theta,\lambda^2)\pi(\ms\theta,\lambda^2|\mb Z).
\end{equation*}
This means that $p_B(\mb Z)$ is simply the posterior expectation of the initial $p$-value: 
\begin{eqnarray*}
p_B(\mb Z) &=& \mathbb E [p_{val}(\ms\theta, \lambda^2, \mb Z) | \mb Z] = \int_{\ms\theta, \lambda^2} p_{val}(\ms\theta, \lambda^2, \mb Z) \pi(\ms\theta,\lambda^2|\mb Z)\dd\ms\theta \dd\lambda^2,
\end{eqnarray*}
which we approximate here by the empirical mean 
$$
\widehat{p_B}(\mb Z) = \frac{1}{M} \sum_m^M p_{val}(\ms\theta_m, \lambda_m^2, \mb Z),$$
as illustrated in Figure~\ref{fig:scatt} (left). Recall that in our case $p_{val}(\ms\theta_m, \lambda_m^2, \mb Z)$ is given by the probability that a chisquare variate with $T$~df exceeds the observed discrepancy~$\mathcal X^2(\mb Z,\ms\theta_m,\lambda_m^2)$. As usual in a Bayesian approach, we deal here with uncertainty on parameters by {\em integrating} over them, as opposed to setting them to estimated values, as it would be done in a frequentist setting. Finally, the resulting Bayesian test mimics the original test by rejecting $\mathcal H_0$ at a user-chosen level $\alpha$ whenever $p_B(\mb Z) \leq\alpha$.

Alternatively, $p_B(\mb Z)$ can be estimated using realisations $(\mb Z_{rep,m},\ms\theta_m,\lambda_m^2)_{1\leq m \leq M}$, where the $(\ms \theta_m,\lambda_m^2)$ are given by the calibration procedure, and:
\begin{equation*}
\mb Z_{rep,m}\underset{i.i.d.}\thicksim\mathcal{N}\big(y_{\ms\theta_m}(.),\lambda_m^2\big).
\end{equation*}
From (\ref{eq:post_p_value}), a second Monte-Carlo estimate of $p_B(\mb Z)$ is then given by:
\begin{eqnarray*}
\widehat{p_B}(\mb Z) &=& \frac{1}{M} \sum_m^M \mb 1_{\displaystyle{\left\{\mathcal X^2(\mb Z_{rep,m},\ms\theta_m,\lambda_m^2)>\mathcal X^2(\mb Z,\ms\theta_m,\lambda_m^2)\right\}}}.
\end{eqnarray*}
This estimator is more general than the previous one, since it requires no evaluation of the conditional $p$-value $p_{val}(\ms\theta, \lambda^2, \mb Z)$, and thus can be used if the latter has no closed form\footnote{This generality comes at a cost, since the Monte-Carlo variance associated with this second estimator is systematically higher than with the first one.}.

Figure \ref{fig:scatt}, right shows the scatterplot of $\mathcal{X}^{2}(\mb Z_{rep},\ms\theta,\lambda^2)$ against $\mathcal{X}^{2}(\mb Z,\ms\theta,\lambda^2)$.
By either method, $p_B$ is found equal to $0.63$, meaning that there is no evidence to reject model (\ref{eq:modelstat}).

\section{Optimal power consumption forecasts}
\label{sec:five}

Most energy contracts are based on a description of the building occupied by customers, accounting for all electrical devices inside the building. This description is used to predict the nominal level of power included in the contract, which can be modified later if the effective consumption is far away from the initial forecast. In all cases, the monthly bill payed by the customer depends on both the type of contract he has, and his effective consumption.

In contrast, it is now common in many other lines of business, such as web access / data services, to propose so-called `unlimited access' contracts, for which the customer pays a fixed fee, based on an initial estimation of his needs. In this section, we investigate how such energy contracts could be designed, based on probabilistic forecasts such as illustrated in Figure~\ref{fig:pred_puiss_moy}. 

In short, the problem can be summarized as that of computing a point estimate of the average electric consumption $\bar{\ms Y}(\ms\theta)$ delivered inside the experimental cell over a certain period of time. The standard approach would suggest computing a central value, such as the mean, median or maximum of the \textit{a posteriori} predictive distribution; see \cite{Berger85}.
However, such choices don't take into account the underlying stakes, which is hardly possible in an industrial context.

The link between statistical estimation and decision under uncertainty in industrial studies has been carefully studied by \cite{Pasanisi2012b}, grounded in Bayesian decision theory (described, e.g., in \cite{Berger85,Bernardo+94,Robert+98,French2000}). We now recall the main ingredients of this approach applied to the field of energetic building modeling. 

Bayesian inference provides a measure of the uncertainty on quantities of interest under the form of a probability distribution. If a point estimate is required, a Bayes estimator should be calculated based on both the probability distribution of the quantity and a \textit{cost function} which assesses the economical consequences of all possible estimation errors, that is, the differences between all candidate point estimates and the unknown true value of the quantity. 

More formally, let $d$ be the energy forecast used to define an energy contract for a new customer. The cost function, denoted by $C(d,\bar{\ms P})$, measures the economical consequences induced by the choice of $d$ instead of the true average consumption $\bar{\ms P}$ (of course unknown in advance). 
Equivalently, the cost function can be replaced by an \textit{utility function} as an assessment of the profit instead of the loss. In this case, the associated Bayes estimate $\hat{d}$ maximizes the average utility:
\begin{equation*}
\hat{d}=\argmax{d}{\int_{\bar{\ms P}}U(d,\bar{\ms P})\pi(\bar{\ms P} | \mb Z)\dd \bar{\ms P}}.
\end{equation*}

For illustration purposes, we now describe a possible utility function that can be used to optimize an `unlimited access' energy contract, assuming that the customer pays a fixed fee $d$. In the case where the effective consumption $\bar{\ms P} $ exceeds $d$, the energy supplier commits to pay the surplus amount, thus ensuring customer satisfaction (and hence loyalty). This first goal alone suggests under-estimating drastically $\bar{\ms P} $, with the risk of inciting customers to waste the energy they haven't payed for. Hence, a company which proposes such a contract should carefully assess its benefits.

To this end, we suggest using the following utility function, based on a linear energy price:
\begin{equation}
\label{Util1}
U(d,\bar{\ms P})=(m\times d)\mathbf{1}_{\{\bar{\ms P}>d\}}+\frac{m\times d}{c(d-\bar{\ms P})+1}\mathbf{1}_{\{\bar{\ms P}\leq d\}},
\end{equation}
where
\begin{itemize}
\item $m$ is the marginal electricity price,
\item $c$ characterizes the probability $1-(c(d-\bar{\ms P})+1)^{-1}$ that the customer breaks the contract, given that he pays more than he has consumed. For small values of $c$, this probability is approximately equal to $c(d-\bar{\ms P})$, hence $c$ can be interpreted as a {\em customer defection rate}, assuming that the number of defecting customers is proportional to the amount of unused energy they have payed for.
\end{itemize}
%

\begin{figure}
\centering
\includegraphics[scale=0.55]{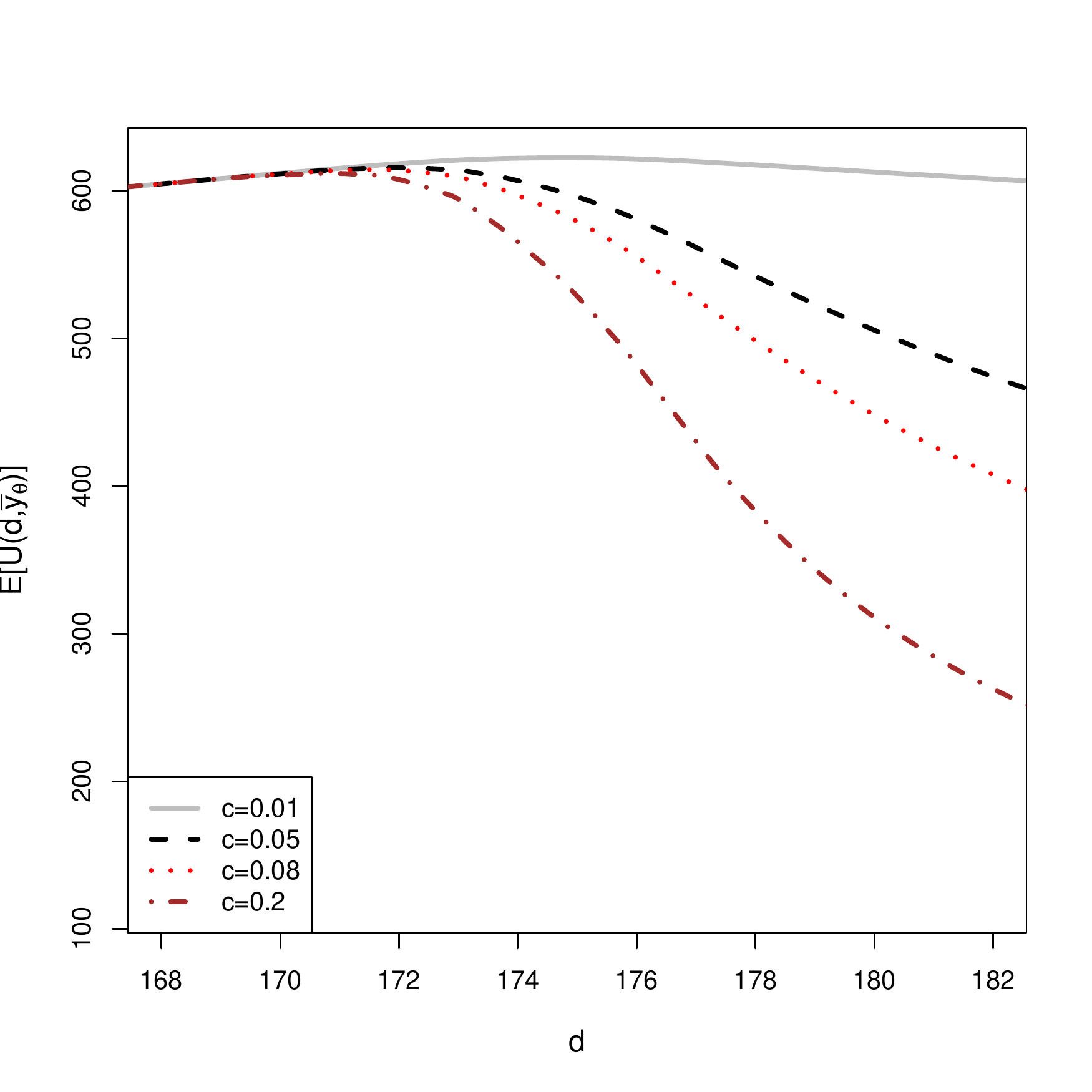}
\caption{\label{fig:average_utility} Expected utilities as functions of the fixed contract fee $d$, for different defection rates $c$.}
\end{figure}

Figure \ref{fig:average_utility} shows the optimal energy forecast $\hat d$ corresponding to several values of the defection rate $c$, given a certain marginal energy price $m$. Note that, due to our linear price assumption, $m$ simply acts as a scaling factor and does not in fact influence the choice of $\hat d$, which is entirely driven by $c$. Accordingly, a small defection rate $c=1\%$ (solid line) leads to $\hat d = 178$, close to the MAP estimate (as seen in Figure~\ref{fig:pred_puiss_moy}), whereas the higher rate of $c=20\%$ (dot-dashes) leads to the much lower optimal value $\hat d = 172$, necessarily yielding lower benefits (the dot-dashed curve has a lower maximum than the solid line curve).

These results show how, with the help of the proposed utility function, we can quantify precisely how, in a highly competitive market, companys must cut down on benefits in order to perserve market share. Obviously, such results are highly sensitive to the tuning of $m$ and $c$ which should be carefully assessed with the help of the trade sector of the company. These could in turn be estimated based on existing data, in which case they would be added to the list of uncertain parameters over which the expectation of the utility is computed.



\section{Discussion}

\label{sec:discussion}

In this paper, we have proposed a practical Bayesian methodology for the calibration and validation of a computer code used to forecast a quantity of interest in an industrial study, used to guide certain strategic decisions for the company, and taking into account the stakes behind these decisions, based on Bayesian decision theory. We illustrated this methodology in a case study concerned with forecasting the energy consumed to heat a building, in view of optimizing an electricity contract including a guarantee of energy supply.

The method presented here is fairly generic, and can in principle be applied to virtually any application domain, as soon as some important decisions need to be taken, based on the behavior of a physical system with uncertain outputs. 

Importantly, note that the three steps of the validation process we propose, including:
\begin{itemize}
\item statistical testing of the code's goodness of fit,
\item parametric uncertainty propagation through the code, 
\item and posterior utility maximization,
\end{itemize}
are entirely based on the output of the Bayesian calibration step, and require no additional code runs. This makes our approach very attractive from a computational perspective, since it requires only minimal efforts once calibration has been done.

Nevertheless, the practical implementation of our approach still depends on the different elements at hand. In the simplified setting presented here, the calculations were fairly standard, but many exciting challenges remain to be met in order to apply the proposed framework to more realistic settings, as discussed in the following.

To begin with, we are currently working on extending the time period over which the code is calibrated, then validated. This raises several issues, even when considering a single cell (rather than a complete building). Indeed, running the code over an extended period increases the computational cost, requiring the use of a metamodel to speed up calculations, such as the dynamic emulator proposed in \cite{Liu2004}. Because data from different periods correspond to different experimental settings, considering different versions of the code (depending on whether the temperature or the power, or both, are considered as outputs of interest), subject to dynamical constraints, constitutes a promising perspective. Another challenge is posed by the albedo. In the paper, this uncertain parameter is assumed constant but it might be more realistic to model it by a time series. This would require more sophisticated calibration procedures.

Furthermore, we have specified some utility functions to commit for an overall consumption forecast to customers according to a new kind of energy contract. The way to construct them remains a considerable challenge, which should only be tackled with the help of the trade sector of the energy supplier. The simplified functions proposed here should be viewed as a guideline for building more realistic ones.

Finally, the Bayesian test proposed here, which mimics the behavior of a classical test, only allows to control the type I (false positive) error rate, but not the type II (false negative) error rate, meaning that we cannot ensure that the code is valid with a given confidence level. Also, we have not dealt with the case where the test rejects the null hypothesis, meaning that a significant discrepancy between model and reality has been detected. 

Both issues can be overcome in the more general context of Bayesian model selection and averaging, as suggested in \cite{Damblin2016}. This would imply comparing the predictions from the statistical model used here, which assumes the code is a perfect representation of reality, with the statistical model in \cite{Koh2001}, which adds a discrepancy term to account for model errors. This would allow to either select the best model for prediction, or combine predictions from both models, weighted by their respective posterior probabilities given the data at hand.

In any case, there is a rising trend for the use of numerical simulations to guide industrial choices, in an increasingly competitive market, under higher and higher safety and regulatory constraints. Hence, the issue of assessing model uncertainty and its impact on decision making is becoming a central question. The methodology we have introduced in this paper is a contribution to address this challenge, but in no way a final answer.




\section{Acknowlegdements}

This work was part of Guillaume Damblin's PhD work at EDF R\&D and AgroParisTech, on the calibration and validation of computer codes. It was partially funded by the French Agence Nationale de la Recherche (ANR), under grant ANR-13-MONU-0005 (project CHORUS).



\begin{thebibliography}{10}

\bibitem{Pasanisi2008}
A.~Pasanisi and J.~Ojalvo.
\newblock Estimation of a quantity of interest in uncertainty analysis: Some
  help from bayesian decision theory.
\newblock {\em {A multi-criteria decision tool to improve the energy efficiency
  of residential buildings}}, 33(1):71--82, 2008.

\bibitem{Eastman2011}
Chuck Eastman, Paul Tiecholz, Rafael Sacks, and Kathleen Liston.
\newblock {\em {CBIM Handbook: A Guide to Building Information Modeling for
  Owners, Managers, Designers, Engineers and Contractors (2nd ed.)}}.
\newblock Hoboken, New Jersey: John Wiley, 2011.

\bibitem{Rysanek2012}
A.~M. Rysanek and R.~Choudhary.
\newblock A decoupled whole-building simulation engine for rapid exhaustive
  search of low-carbon and low-energy building refurbishment options.
\newblock {\em Building and Environment}, 50:21--33, 2012.

\bibitem{Tian2011}
W~Tian and R~Choudhary.
\newblock {Energy use of buildings at urban scale: A case study of london
  school buildings}.
\newblock In {\em Proceedings of Building Simulation 2011: 12th Conference of
  International Building Performance Simulation Association}, pages 1702--1709,
  2011.

\bibitem{Salt00}
A~Saltelli, K.~Chan, and E.M. Scott.
\newblock {\em Sensitivity Analysis}.
\newblock Wiley, New York, 2000.

\bibitem{Heo2012}
Y.~Heo, R.~Choudharyb, and Augenbroea G.A.
\newblock {Calibration of Building Energy Models for Retrofit Analysis under
  Uncertaintyc}.
\newblock {\em Energy and Buildings}, 47:550--560, 2012.

\bibitem{Rivalin2016}
Lisa Rivalin.
\newblock {\em {Vers une démarche de garantie des consommations énergétiques
  dans les bâtiments neufs : Méthodes d’évaluation des incertitudes
  associées à la simulation thermique dynamique dans le processus de
  conception et de réalisation}}.
\newblock PhD thesis, Ecole doctorale n.432 : Sciences des m\'etiers de
  l'ing\'enieur, 2016.

\bibitem{Pasanisi2012a}
A.~Pasanisi and A.~Dutfoy.
\newblock {An Industrial Viewpoint on Uncertainty Quantification in Simulation:
  Stakes, Methods, Tools, Examples}.
\newblock In A.M. Dienstfrey and R.F. Boisvert, editors, {\em Uncertainty
  Quantification in Scientific Computing}, pages 27--45. Springer, 2012.

\bibitem{Baudin2016}
M.~Baudin, , A.~Dutfoy, B.~Iooss, and A-L. Popelin.
\newblock {Open TURNS: An industrial software for uncertainty quantification in
  simulation}.
\newblock In {\em Springer Handbook on Uncertainty Quantification}. Springer,
  2016.

\bibitem{Sacks89}
J.~Sacks, Welch. W.J., T.J. Mitchell, and H.P Wynn.
\newblock Design and analysis of computer experiments.
\newblock {\em Technometrics}, 31:41--47, 1989.

\bibitem{Pasanisi2016}
A.~Pasanisi, A.~Koch, M.~Peter, and B.~Boutaud.
\newblock { Smart and Sustainable Cities: A Systemic Approach }.
\newblock In {\em ENBIS-2016 Conference, Sheffield}, 2016.

\bibitem{Mirakyan2015}
A.~Mirakyan, A.~Nichersu, A.~Pasanisi, M.~Saed, N.~Schweiger, M.~Sipowicz, and
  J.~Wendel.
\newblock { Applied Statistics in Support of Cities Simulation: Some Examples
  and Perspectives }.
\newblock In {\em ENBIS-2015 Conference, Prague}, 2015.

\bibitem{Blin2015}
D.~Blin, F.~Casciani, P.~Imbert, B.~Mousseau, A.~Pasanisi, P.~Terrien, and
  P.~Viejo.
\newblock { A software platform to help Singapore to build a more smart and
  sustainable city}.
\newblock In {\em Submission SFT}, 2015.

\bibitem{Koh2001}
M.~Kennedy and A.~O'Hagan.
\newblock Bayesian calibration of computer models.
\newblock {\em Journal of the Royal Statistical Society, Series B,
  Methodological}, 63:425--464, 2001.

\bibitem{Damblin2015}
Guillaume Damblin.
\newblock {\em Statistical Contributions to calibration and validation of
  compute codes}.
\newblock PhD thesis, École doctorale n.581 : Agriculture, alimentation,
  biologie, environnement et santé, 2015.

\bibitem{Bayarri2007}
M.~J. Bayarri, J.~O. Berger, P.~R. Sacks, J.~A. Cafeo, J.~Cavendish, C.-H. Lin,
  and J.~Tu.
\newblock A framework for validation of computer models.
\newblock {\em Technometrics}, 49:138--154, 2007.

\bibitem{Bontemps2015}
St\'ephanie Bontemps.
\newblock {\em {Empirical validation of models: application to low-energy
  buildings (in french)}}.
\newblock PhD thesis, Doctorale school n.432 : Sciences des m\'etiers de
  l'ing\'enieur, 2015.

\bibitem{Roy11}
C.J. Roy and W.L. Oberkampf.
\newblock A comprehensive framework for verification, validation and
  uncertainty quantification in scientific computing.
\newblock {\em Comput. Methods Appl. Mech. Engrf}, 200:2131--2144, 2011.

\bibitem{Spitz}
C.~Spitz.
\newblock {\em Analyse de la fiabilité des outils de simulation et des
  incertitudes de métrologie appliquée à l'efficacité énergétique des
  bâtiments}.
\newblock PhD thesis, Université de Grenoble, 2012.

\bibitem{Pasanisi2012b}
A.~Pasanisi, M.~Keller, and E.~Parent.
\newblock Estimation of a quantity of interest in uncertainty analysis: Some
  help from bayesian decision theory.
\newblock {\em Reliability Engineering and System Safety}, 100:93--101, 2012.

\bibitem{Bernardo+94}
J.~M. Bernardo and A.~F.~M. Smith.
\newblock {\em {B}ayesian Theory}.
\newblock Wiley, London, 1 edition, 1994.

\bibitem{Cam06}
K.~Campbell.
\newblock Statistical calibrations of computer simulations.
\newblock {\em Reliability Engineering and System Safety}, 91:1358--1363, 2006.

\bibitem{Cox2001}
D.~Cox, J.~Park, and E.~Clifford.
\newblock {A statistical method for tuning a computer code to a data base}.
\newblock {\em Computational Statistics and Data Analysis}, 37:77--92, 2001.

\bibitem{Robert+98}
C.P. Robert and G.~Casella.
\newblock {\em Monte Carlo Statistical Methods}.
\newblock Springer-Verlag, 1998.

\bibitem{French2000}
S.~French and D.R. Insua.
\newblock {\em {Statistical Decision Theory}}.
\newblock Wiley: Kendall's Library of Statistics 9, 2000.

\bibitem{Plessis2014}
G.~Plessis, A.~Kaemmerlin, and A.~Lindsay.
\newblock { BuildSysPro: a Modelica library for modelling buildings and energy
  systems }.
\newblock In {\em Proceedings of the 10th International ModelicaConference,
  Lund, Sweden}, 2014.

\bibitem{Elmqvist1978}
Hilding Elmqvist.
\newblock {\em {A Structured Model Language for Large Continuous Systems}}.
\newblock PhD thesis, Department of Automatic Control, Lund University,
  Sweden., 1978.

\bibitem{Bontemps2013}
S.~Bontemps, A.~Kaemmerlen, R.~Le-Berre, and L.~Mora.
\newblock La fiabilité d'outils de simulation thermique dynamique dans le
  contexte des bâtiments basse consommation.
\newblock {\em Submission SFT}, 2013.

\bibitem{AIAA}
AIAA.
\newblock Guide for the verification and validation of computational fluid
  dynamics simulations.
\newblock {\em American Institute of Aeronautics and Astronautics}, 1998.

\bibitem{Roache1998}
P.J. Roache.
\newblock Verification of codes and calculations.
\newblock {\em AIAA Journal}, 36:696--702, 1998.

\bibitem{Gelman1996}
A.~Gelman, X.L Meng, and H.~Stern.
\newblock Posterior predictive assessment of model fitness via realized
  discrepancies.
\newblock {\em Statistica Sinica}, 6:733--807, 1996.

\bibitem{Efron1981}
B.~Efron.
\newblock {Nonparametric Estimates of Standard Error: The Jackknife, the
  Bootstrap and Other Methods}.
\newblock {\em Biometrika}, 68(3):589--599, 1981.

\bibitem{Vandervaart2000}
A.~W. van~der Vaart.
\newblock {\em Asymptotic Statistics (Cambridge Series in Statistical and
  Probabilistic Mathematics)}.
\newblock {Cambridge University Press}, June 2000.

\bibitem{Berger85}
James~O. Berger.
\newblock {\em Statistical Decision Theory and Bayesian Analysis}.
\newblock Springer, 2nd edition, 1985.

\bibitem{Liu2004}
F.~Liu and M.~West.
\newblock A dynamic modelling strategy for bayesian computer model emulation.
\newblock {\em Bayesian Analysis}, 1(1), 2004.

\bibitem{Damblin2016}
G.~Damblin, M.~Keller, P.~Barbillon, A.~Pasanisi, and E.~Parent.
\newblock {Bayesian Model Selection for the Validation of Computer Codes}.
\newblock {\em QREI}, 32(6 (Special Issue: The ENBIS-15 Quality and Reliability
  Engineering International)):2043--2054, 2016.

\end{thebibliography}

\end{document}